\documentclass[sigconf]{acmart}

\setcopyright{none}
\usepackage[T1]{fontenc}
\usepackage{graphicx}
\usepackage{amsmath}
\usepackage{hyperref}
\usepackage{lipsum}
\usepackage[most]{tcolorbox}
\usepackage{subcaption}
\usepackage{subfiles}
\usepackage{dblfloatfix} % enables [b] for figure*
\usepackage{doi}

\usepackage{tikz}
\usetikzlibrary{arrows.meta, calc, backgrounds, fit}
\usepackage{pgffor}
\usepackage{xstring}

\definecolor{environmentbg}{RGB}{8,230,0} 
\definecolor{datapipelinebg}{RGB}{71,179,255}
\definecolor{modelinferencebg}{RGB}{56,112,232}
\definecolor{applicationbg}{RGB}{100,1,247}  

\definecolor{naturalColor}{RGB}{8,230,0}
\definecolor{featureColor}{RGB}{71,179,255}
\definecolor{inferenceColor}{RGB}{56,112,232}
\definecolor{applicationColor}{RGB}{107,47,247}

\tcbset{
  subtlebox/.style={
    colframe=black!5,
    boxrule=0.3pt,
    arc=2pt,
    left=6pt,
    right=6pt,
    top=4pt,
    bottom=4pt,
    enhanced,
  }
}
\begin{document}

\title{Tracing Distribution Shifts with Causal System Maps}

% \title{Explaining Shifts in Machine Learning Systems with Causal Maps}

% \title{Explaining Distribution Shifts with Causal System Maps}
% \title{Explaining Shifts with Machine Learning System Maps}
% \title{Tracing Distribution Shifts with Machine Learning System Maps}

\author{Joran Leest}
\affiliation{
  \institution{Vrije Universiteit Amsterdam}
  \country{The Netherlands}
}
\email{j.g.leest@vu.nl}

\author{Ilias Gerostathopoulos}
\affiliation{
  \institution{Vrije Universiteit Amsterdam}
  \country{The Netherlands}
}
\email{i.g.gerostathopoulos@vu.nl}

\author{Claudia Raibulet}
\affiliation{
  \institution{Universita' degli Studi di Milano-Bicocca}
  \country{Italy}
}
\email{claudia.raibulet@unimib.it}

\author{Patricia Lago}
\affiliation{
  \institution{Vrije Universiteit Amsterdam}
  \country{The Netherlands}
}
\email{p.lago@vu}

% \institute{
%   Vrije Universiteit Amsterdam \\
%   \email{\{j.leest, i.gerostathopoulos, c.raibulet\}@vu.nl}
% }

\begin{abstract}
Monitoring machine learning (ML) systems is hard, with standard practice focusing on detecting distribution shifts rather than their causes. Root-cause analysis often relies on manual tracing to determine whether a shift is caused by software faults, data-quality issues, or natural change. We propose \emph{ML System Maps} -- causal maps that, through layered views, make explicit the propagation paths between the environment and the ML system’s internals, enabling systematic attribution of distribution shifts. We outline the approach and a research agenda for its development and evaluation.
\end{abstract}

\maketitle

\section{Introduction}\label{intro}

Machine learning (ML) systems contain data-driven components whose logic cannot be fully inspected or understood \cite{rudin2022interpretable}. Their behavior depends on many upstream factors -- both internal data processing (feature engineering, transformations) and external data-generating processes \cite{leest2025tea,shankars2024we,breck2017ml,huyen2022}. Distribution shifts in any of these dependencies can cause the model to stop generalizing, deteriorating the decisions it drives \cite{lu2018,breck2017ml}.

In operational ML systems, distributional shifts serve as the primary monitoring signal, indicating that \textit{something} changed but not \textit{why} \cite{shankars2024we,shergadwalamn2022a, sculley2015hidden,huyen2022,xuz2023alertiger}. The root cause might be data quality issues \cite{breck2019data, ehrlingerl2019a}, deployment bugs \cite{shankars2024we, huyen2022}, changing user behavior \cite{cloudera2021, vzliobaite2015overview}, or even the system's own actions \cite{she2025fairsense, feng2024monitoring}. Each requires different responses -- e.g., reverting a pipeline change, or model retraining -- yet shifts appear as undifferentiated symptoms.

Effective monitoring thus requires solving an \textit{attribution} problem: tracing an observed shift from symptom to source -- determining whether a shift originates locally, is inherited from upstream dependencies, or reflects a natural change \cite{protschky2025gets,shankar2023automatic,shankars2022towards}. This often requires extensive manual investigation; a process that can be understood as answering three attribution questions (AQs).

\textbf{AQ1: \textit{Which subsystem is responsible for the shift, and which team should it be escalated to?}} ML systems span organizational boundaries -- data teams maintain pipelines, ML teams manage models, business teams consume predictions \cite{kreuzberger2023machine}. Tight coupling causes shifts to propagate downstream, creating symptoms in multiple subsystems \cite{breck2017ml}; attribution first requires determining whether a shift originates locally or was inherited upstream.

\textbf{AQ2: \textit{Which component within the identified subsystem could have introduced the shift?}} Once the responsible subsystem is identified, the owning team must first confirm whether the shift originated locally and, if so, isolate potential faults (e.g., a stale table, code change, or faulty configuration). This is difficult because subsystems comprise many interdependent components \cite{shankars2022towards}.

\textbf{AQ3: \textit{If the shift is not caused internally, could a natural change in the underlying process explain it?}} Not every shift indicates a problem -- ML systems model external processes that continuously change \cite{cobbo2022contextaware, cloudera2021}. If a shift stems from changes in external processes, we can rule out internal faults (i.e., ``explain it away'' \cite{wellman1993explaining}) -- e.g., a shift might simply match a recurring pattern. Even in this case, the exact source needs to be understood to separate benign shifts from those that invalidate modeling assumptions \cite{webb2016characterizing, huyen2022}.

We envision a systematic approach to answering these questions automatically through \textit{ML System Maps} -- a hierarchically structured representation of the cause-and-effect relationships within an ML system and its environment. ML System Maps build on the conceptual framework of \cite{leest2025tea} and draw on causal inference methods \cite{quintas2024multiply, budhathokik2021why, paleyes2023dataflow, singal2021flow}. They enable systematic attribution through three coordinated views, each answering one AQ: the \textit{ML System View} (AQ1), \textit{Subsystem View} (AQ2), and \textit{Environment View} (AQ3). We develop this vision by first characterizing operational ML systems and their structure (Section~\ref{system_characterization}), then we introduce our method (Section~\ref{sec:maps}), illustrating its attribution capabilities (Section~\ref{explaining_drifts}), and discuss practical considerations and a research agenda (Section~\ref{sec:discussion}).

\section{The Structure of an ML system} \label{system_characterization}

To analyze how changes propagate through an ML system, we first need to understand how ML systems are structured. In this section, we characterize the structure of ML systems and establish key concepts for systematic representation. We distinguish the \textbf{internal ML system} (managed subsystems within organizational control) from the \textbf{environment} (data-generating  processes that drive the system's behavior) \cite{leest2025tea}. Throughout the paper, we use a \textit{running example} to illustrate the concepts and our approach;

\begin{tcolorbox}[colback=gray!10, boxrule=0.5pt, colframe=black!100, arc=1mm]
\textbf{Running example.} \textit{A company uses an ML system to reduce customer churn through targeted promotions. The system processes user features through a churn risk model and a promotion-ranking model. These predictions inform two sequential decisions: identifying which at-risk customers to contact, and selecting which promotional offers to send out.}
\end{tcolorbox}

\subsection{Internal ML System}\label{internal_subsystems}

The internal ML system comprises three subsystems through which data flows \cite{leest2025tea}: a \textbf{data pipeline} produces features for \textbf{model serving}, which generates predictions used by an \textbf{application} to make business decisions. Each subsystem itself can be understood as a data flow, has a canonical responsible stakeholder, and can introduce distribution shifts:

\begin{tcolorbox}[subtlebox, breakable, colback=datapipelinebg!8]
\textbf{Data Pipeline.} Maintained by \textit{data engineers}, this subsystem involves the extract-transform-load (ETL) processes that ingest source data, apply transformations through aggregations and joins, and load the transformed data into feature tables. Tools such as dbt \cite{dbt_docs} represent pipelines as directed acyclic graphs (DAGs), with transformations and artefacts as nodes and dependencies as edges. In processing the data, distribution shifts can be introduced through scheduled updates (e.g., schema changes) or faults (e.g., failed jobs) -- causing data quality issues that propagate downstream \cite{shankars2024we}.

\textbf{Example}: \textit{The pipeline parses raw user activity logs into daily activity counts, then computes rolling averages and other aggregations to produce features for the churn model. Parsing errors in the raw logs lead to under-represented activity, while failed jobs produce stale features -- both introducing data quality issues.}
\end{tcolorbox}

\vspace{-0.3cm}

\begin{tcolorbox}[subtlebox, breakable, colback=modelinferencebg!8]
\textbf{Model Serving.} Managed by \textit{ML engineers}, this subsystem involves the serving pipeline that loads features from tables, applies pre-processing, runs inference through ML models, and produces predictions. Frameworks like TensorFlow Extended~\cite{tfx} represent these serving flows as pipeline graphs, tracking model versions and serving configurations. Distribution shifts can arise from intentional changes or faults, potentially compromising predictions and propagating to dependent models (often managed by different teams) \cite{sculley2015hidden,breck2017ml}.

\textbf{Example}: \textit{The serving pipeline loads demographic and activity features, runs inference through the churn risk model to generate predictions, which are then consumed by the promotion-ranking model. Routing traffic to different model versions (e.g., during experiments) or timeouts in fetching features can shift churn predictions and propagate to the dependent model.}
\end{tcolorbox}

\vspace{-0.3cm}

\begin{tcolorbox}[subtlebox, breakable, colback=applicationbg!8]
\textbf{Application.} Managed by \textit{business stakeholders}, applications combine predictions with business rules to produce business decisions. These range from simple rule-based logic (e.g., threshold-based decisions) to complex workflows (e.g., Salesforce \cite{salesforce_marketing_automation} or Zapier \cite{zapier_workflows} automations). Applications can cause distribution shift via selection effects, where decisions drive actions that determine which cases downstream models observe -- and the data they are trained on \cite{fengj2024designing,she2025fairsense}.

\textbf{Example}: \textit{The retention team uses churn predictions combined with business rules (budget constraints, customer priority) to make outreach decisions. A separate promotions team then selects which specific offer to send based on rankings from the promotion model. A change in outreach policy risks exposing the promotion-ranking model to a shifted user population, potentially beyond what it can generalize to.}
\end{tcolorbox}

Having described these subsystems as data flows, we leverage the insight of \cite{paleyes2023dataflow} that such flows can be translated into a cause-and-effect diagram (or causal graph) by representing the outputs of processes as causally connected nodes. The principle is straightforward: if a downstream variable is derived from upstream variables, it is causally determined by them. For example, at the ML system level, features cause predictions, which cause decisions \cite{leest2025tea}.

\subsection{Environment}\label{environment}

ML systems are data-driven; their behavior is fundamentally \textit{conditioned on} the environment they observe and act upon, not just internally managed processes. Without  accounting for these changing external data-generating processes, shift attribution  remains incomplete. Following \cite{leest2025tea}, we treat the environment as an integral system to consider jointly with the internal ML system.

\begin{tcolorbox}[subtlebox, colback=environmentbg!8]
\textbf{Environment.} The real world that generates data for the ML system and is affected by its decisions. Changes in the environment arise from trends and seasonality (e.g., holiday patterns) or from unexpected events (i.e., exogenous shocks) like logistics disruptions or service outages \cite{cobbo2022contextaware,webb2016characterizing}. Since these changes occur before measurement, they propagate through the ML system (from data processing to application) and can resemble internal faults \cite{shankars2022towards,xuz2023alertiger}. While teams do not control the environment, \textit{data scientists} typically track trends and events to ensure modeling assumptions remain valid \cite{zhang2020data}.

\textbf{Example}: \textit{The environment represents the customer churn 
process -- how behavior leads to churn outcomes -- which the ML system observes and influences through promotions.}
\end{tcolorbox}

Like the internal ML system, the environment has a cause-and-effect structure, such as user demographics shaping behavior. While reasoning over this structure is common (though often left implicit) in model development \cite{lamsaf2025causality}, explicitly formalizing this structure enables tracing shifts to their source -- e.g., a shift in observed user demographics explaining changes in their activity patterns \cite{budhathokik2021why}.

% For example, if we observe a shift in user behavior, and demographics causally affect behavior, then observing a concurrent demographic shift can explain the behavior change as a population shift.

% However, such explanations require first solving an attribution problem: determining whether an observed distribution shift originates in the real-world process itself or in how the internal system measures and transforms that process -- a problem we address in Section \ref{attribution}.

\section{ML System Maps} \label{sec:maps}

Building on this causal framing, we introduce our approach \textit{ML System Maps}: a method for modeling the causal structure of a system, and using it for explaining distribution shifts.
The causal structure is specified by a \textit{notation} (Section~\ref{notation}) and a set of \textit{views} (Section~\ref{views}) organized to systematically answer the three attribution questions mentioned in Section~\ref{intro}. Figure~\ref{fig:system_map} shows an \textit{ML System Map} for the customer retention example, expressed in the notation and views. 

% We illustrate how our approach enables systematic attribution in Section \ref{explaining_drifts}.

\usetikzlibrary{arrows.meta,calc}

% ============================================
% NODE LABELS (appearance + vertical offset)
% ============================================
\providecommand{\nodelabeloffset}{0.2cm} % distance below each node
\newcommand{\labelFontSize}{\footnotesize}

% Put near your other parameters
\providecommand{\nodelabelxpad}{2pt}   % left/right padding
\providecommand{\nodelabelypad}{1pt}   % top/bottom padding
\providecommand{\nodelabelminw}{0pt}   % optional minimum width
\providecommand{\nodelabelminh}{0pt}   % optional minimum height
\providecommand{\nodelabelcorners}{1pt}% corner radius
\providecommand{\nodelabelfill}{black!4} % background shade
\providecommand{\nodelabelborder}{none}   % or e.g., black!30
\providecommand{\nodelabelopacity}{1}     % 0..1

\tikzset{
  nodelabel/.style={
    rectangle,
    fill=\nodelabelfill,
    rounded corners=\nodelabelcorners,
    inner xsep=\nodelabelxpad,
    inner ysep=\nodelabelypad,
    minimum width=\nodelabelminw,
    minimum height=\nodelabelminh,
    draw=\nodelabelborder,
    font=\labelFontSize,
    text opacity=1,
    fill opacity=\nodelabelopacity
  }
}
% ============================================
% CONTROLS (Parameters)
% ============================================
\providecommand{\panelhsep}{0.03\textwidth}       % Horizontal separation between panels
\providecommand{\panelvsep}{0.5cm}                % Vertical separation between rows
\providecommand{\panelheight}{3.3cm}                % Height of all panels

% ============================================
% GRAPH POSITION OFFSETS (Individual control for each graph)
% ============================================
% Top-left graph offsets
\providecommand{\tlGraphXOffset}{0cm}  % Horizontal offset (positive = right, negative = left)
\providecommand{\tlGraphYOffset}{-0.1cm}  % Vertical offset (positive = up, negative = down)

% Top-right graph offsets
\providecommand{\trGraphXOffset}{0cm}
\providecommand{\trGraphYOffset}{-0.1cm}

% Bottom-left graph offsets
\providecommand{\blGraphXOffset}{0cm}
\providecommand{\blGraphYOffset}{-0.1cm}

% Bottom-middle graph offsets
\providecommand{\bmGraphXOffset}{0cm}
\providecommand{\bmGraphYOffset}{-0.1cm}

% Bottom-right graph offsets
\providecommand{\brGraphXOffset}{0cm}
\providecommand{\brGraphYOffset}{-0.1cm}

% ============================================
% TEXT BOX PARAMETERS (Title and Description for each panel)
% ============================================
% Top-left text
\providecommand{\tlTitle}{Environment View}
\providecommand{\tlDesc}{}
\providecommand{\tlTextXOffset}{0cm}  % Horizontal offset for text box
\providecommand{\tlTextYOffset}{0.1cm}  % Vertical offset for text box

% Top-right text
\providecommand{\trTitle}{ML System View}
\providecommand{\trDesc}{}
\providecommand{\trTextXOffset}{0cm}
\providecommand{\trTextYOffset}{0.1cm}

% Bottom-left text
\providecommand{\blTitle}{Data Pipeline View}
\providecommand{\blDesc}{}
\providecommand{\blTextXOffset}{-0.8cm}
\providecommand{\blTextYOffset}{0.1cm}

% Bottom-middle text
\providecommand{\bmTitle}{Model Serving View}
\providecommand{\bmDesc}{}
\providecommand{\bmTextXOffset}{-0.8cm}
\providecommand{\bmTextYOffset}{0.1cm}

% Bottom-right text
\providecommand{\brTitle}{Application View}
\providecommand{\brDesc}{}
\providecommand{\brTextXOffset}{-0.8cm}
\providecommand{\brTextYOffset}{0.1cm}

% Text styling parameters
\providecommand{\titleFontSize}{\large}
\providecommand{\descFontSize}{\small}
\providecommand{\textBoxWidth}{0.8}  % Fraction of panel width for text box

\tikzset{
  node arrow/.style={
    draw,                % ensures it draws even if you use \path
    black,
    line width=0.1pt,
    -{Triangle[length=1.4mm, width=1.4mm]} % longer, not wider
  }
}

% Panel styling
\providecommand{\panellinewidth}{0.1pt}             % Panel border thickness
\providecommand{\panelrounded}{3pt}               % Panel corner radius
\providecommand{\panelfillcolor}{black!4}           % Panel fill color
\providecommand{\panelbordercolor}{white}         % Panel border color

% Node parameters
\providecommand{\nodesize}{0.5cm}                 % Node diameter
\providecommand{\nodesep}{3.089cm}                  % Horizontal separation between nodes (xsep)
\providecommand{\nodevsep}{1.44cm}                 % Vertical separation between node rows (ysep)

% Calculate panel widths to fit exactly within \textwidth
% Top row: 2 panels + 1 gap = \textwidth
\pgfmathsetlengthmacro{\panelwidth}{(\textwidth - \panelhsep)/2}
% Bottom row: 3 panels + 2 gaps = \textwidth
\pgfmathsetlengthmacro{\panelwidthsmall}{(\textwidth - 2*\panelhsep)/3}

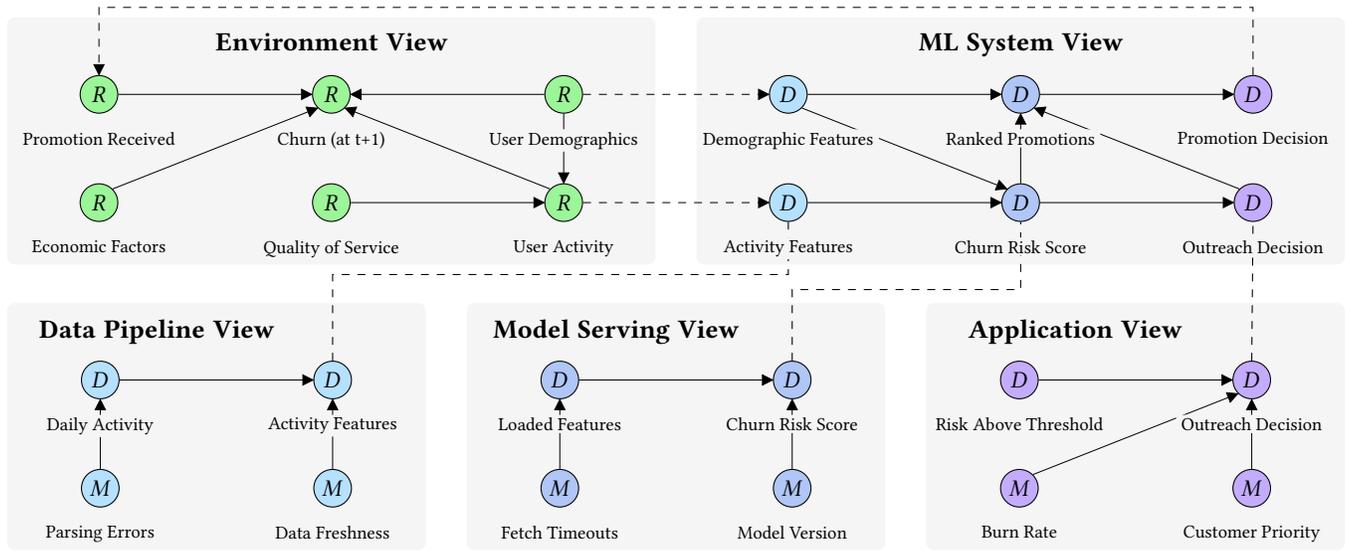
\begin{figure*}[htpb]

\vspace{-0.6cm}

% \noindent\rule{\textwidth}{0.4pt}
% \vspace{-0.01cm} % Add some space between line and figure

\centering
\noindent\begin{tikzpicture}[scale=1]
\clip (0, -\panelheight - \panelvsep - 1cm) rectangle (\textwidth, \panelheight + 1cm);
% ============================================
% TOP ROW: 2 Large Panels
% ============================================

% Top-left panel
\draw[line width=\panellinewidth, draw=\panelbordercolor, fill=\panelfillcolor, rounded corners=\panelrounded] 
  (0, 0) rectangle ++(\panelwidth, \panelheight);

% Top-left text box
\begin{scope}[shift={(0.5*\panelwidth + \tlTextXOffset, \panelheight - \tlTextYOffset)}]
  \node[anchor=north, text width=\textBoxWidth*\panelwidth, align=center] (tltitle) at (0,0) {\titleFontSize\bfseries\tlTitle};
  \node[anchor=north, text width=\textBoxWidth*\panelwidth, align=center] (tldesc) at (0,-0.4cm) {\descFontSize\tlDesc};
\end{scope}

% Draw 2×3 grid in top-left panel with position offsets
\begin{scope}[shift={(0.5*\panelwidth + \tlGraphXOffset, 0.5*\panelheight + \tlGraphYOffset)}]
  % Row 1: nodes 1, 2, 3
  \node[circle, draw, fill=naturalColor!40, minimum size=\nodesize, inner sep=0pt] (tl1) at (-\nodesep, 0.5*\nodevsep) {$R$};
  \node[circle, draw, fill=naturalColor!40, minimum size=\nodesize, inner sep=0pt] (tl2) at (0, 0.5*\nodevsep) {$R$};
  \node[circle, draw, fill=naturalColor!40, minimum size=\nodesize, inner sep=0pt] (tl3) at (\nodesep, 0.5*\nodevsep) {$R$};
  Row 2: nodes 4, 5, 6
  \node[circle, draw, fill=naturalColor!40, minimum size=\nodesize, inner sep=0pt] (tl4) at (-\nodesep, -0.5*\nodevsep) {$R$};
  \node[circle, draw, fill=naturalColor!40, minimum size=\nodesize, inner sep=0pt] (tl5) at (0, -0.5*\nodevsep) {$R$};
  \node[circle, draw, fill=naturalColor!40, minimum size=\nodesize, inner sep=0pt] (tl6) at (\nodesep, -0.5*\nodevsep) {$R$};

  \draw[node arrow] (tl1.east) -- (tl2.west);
  \draw[node arrow] (tl4.north east) -- (tl2.south west);
  \draw[node arrow] (tl3.west) -- (tl2.east);
  \draw[node arrow] (tl3.south) -- (tl6.north);
  \draw[node arrow] (tl6.north west) -- (tl2.south east);
  \draw[node arrow] (tl5.east) -- (tl6.west);
\end{scope}

% Top-right panel
\draw[line width=\panellinewidth, draw=\panelbordercolor, fill=\panelfillcolor, rounded corners=\panelrounded] 
  (\panelwidth + \panelhsep, 0) rectangle ++(\panelwidth, \panelheight);

% Top-right text box
\begin{scope}[shift={(\panelwidth + \panelhsep + 0.5*\panelwidth + \trTextXOffset, \panelheight - \trTextYOffset)}]
  \node[anchor=north, text width=\textBoxWidth*\panelwidth, align=center] (trtitle) at (0,0) {\titleFontSize\bfseries\trTitle};
  \node[anchor=north, text width=\textBoxWidth*\panelwidth, align=center] (trdesc) at (0,-0.4cm) {\descFontSize\trDesc};
\end{scope}

% Draw 2×3 grid in top-right panel with position offsets
\begin{scope}[shift={(\panelwidth + \panelhsep + 0.5*\panelwidth + \trGraphXOffset, 0.5*\panelheight + \trGraphYOffset)}]
  % Row 1: nodes 1, 2, 3
  \node[circle, draw, fill=featureColor!40, minimum size=\nodesize, inner sep=0pt] (tr1) at (-\nodesep, 0.5*\nodevsep) {$D$};
  \node[circle, draw, fill=inferenceColor!40, minimum size=\nodesize, inner sep=0pt] (tr2) at (0, 0.5*\nodevsep) {$D$};
  \node[circle, draw, fill=applicationColor!40, minimum size=\nodesize, inner sep=0pt] (tr3) at (\nodesep, 0.5*\nodevsep) {$D$};
  % Row 2: nodes 4, 5, 6
  \node[circle, draw, fill=featureColor!40, minimum size=\nodesize, inner sep=0pt] (tr4) at (-\nodesep, -0.5*\nodevsep) {$D$};
  \node[circle, draw, fill=inferenceColor!40, minimum size=\nodesize, inner sep=0pt] (tr5) at (0, -0.5*\nodevsep) {$D$};
  \node[circle, draw, fill=applicationColor!40, minimum size=\nodesize, inner sep=0pt] (tr6) at (\nodesep, -0.5*\nodevsep) {$D$};

  \draw[node arrow] (tr1.east) -- (tr2.west);
  \draw[node arrow] (tr4.east) -- (tr5.west);
  \draw[node arrow] (tr5.north) -- (tr2.south);
  \draw[node arrow] (tr2.east) -- (tr3.west);
  \draw[node arrow] (tr5.east) -- (tr6.west);
  \draw[node arrow] (tr6.north west) -- (tr2.south east);
  \draw[node arrow] (tr1.south east) -- (tr5.north west);
\end{scope}

% ============================================
% BOTTOM ROW: 3 Small Panels
% ============================================

% Bottom-left panel
\draw[line width=\panellinewidth, draw=\panelbordercolor, fill=\panelfillcolor, rounded corners=\panelrounded] 
  (0, -\panelheight - \panelvsep) rectangle ++(\panelwidthsmall, \panelheight);

% Bottom-left text box
\begin{scope}[shift={(0.5*\panelwidthsmall + \blTextXOffset, -\panelvsep - \blTextYOffset)}]
  \node[anchor=north, text width=\textBoxWidth*\panelwidthsmall, align=center] (bltitle) at (0,0) {\titleFontSize\bfseries\blTitle};
  \node[anchor=north, text width=\textBoxWidth*\panelwidthsmall, align=center] (bldesc) at (0,-0.4cm) {\descFontSize\blDesc};
\end{scope}

% Draw 2×2 grid in bottom-left panel with position offsets
\begin{scope}[shift={(0.5*\panelwidthsmall + \blGraphXOffset, -\panelheight - \panelvsep + 0.5*\panelheight + \blGraphYOffset)}]
  % Row 1: nodes 1, 2
  \node[circle, draw, fill=featureColor!40, minimum size=\nodesize, inner sep=0pt] (bl1) at (-0.5*\nodesep, 0.5*\nodevsep) {$D$};
  \node[circle, draw, fill=featureColor!40, minimum size=\nodesize, inner sep=0pt] (bl2) at (0.5*\nodesep, 0.5*\nodevsep) {$D$};
  % Row 2: nodes 3, 4
  \node[circle, draw, fill=featureColor!40, minimum size=\nodesize, inner sep=0pt] (bl3) at (-0.5*\nodesep, -0.5*\nodevsep) {$M$};
  \node[circle, draw, fill=featureColor!40, minimum size=\nodesize, inner sep=0pt] (bl4) at (0.5*\nodesep, -0.5*\nodevsep) {$M$};
  
  \draw[node arrow] (bl1.east) -- (bl2.west);
  \draw[node arrow] (bl4.north) -- (bl2.south);
  \draw[node arrow] (bl3.north) -- (bl1.south);
\end{scope}

% Bottom-middle panel
\draw[line width=\panellinewidth, draw=\panelbordercolor, fill=\panelfillcolor, rounded corners=\panelrounded] 
  (\panelwidthsmall + \panelhsep, -\panelheight - \panelvsep) 
  rectangle ++(\panelwidthsmall, \panelheight);

% Bottom-middle text box
\begin{scope}[shift={(\panelwidthsmall + \panelhsep + 0.5*\panelwidthsmall + \bmTextXOffset, -\panelvsep - \bmTextYOffset)}]
  \node[anchor=north, text width=\textBoxWidth*\panelwidthsmall, align=center] (bmtitle) at (0,0) {\titleFontSize\bfseries\bmTitle};
  \node[anchor=north, text width=\textBoxWidth*\panelwidthsmall, align=center] (bmdesc) at (0,-0.4cm) {\descFontSize\bmDesc};
\end{scope}

% Draw 2×2 grid in bottom-middle panel with position offsets
\begin{scope}[shift={(\panelwidthsmall + \panelhsep + 0.5*\panelwidthsmall + \bmGraphXOffset, -\panelheight - \panelvsep + 0.5*\panelheight + \bmGraphYOffset)}]
  % Row 1: nodes 1, 2
  \node[circle, draw, fill=inferenceColor!40, minimum size=\nodesize, inner sep=0pt] (bm1) at (-0.5*\nodesep, 0.5*\nodevsep) {$D$};
  \node[circle, draw, fill=inferenceColor!40, minimum size=\nodesize, inner sep=0pt] (bm2) at (0.5*\nodesep, 0.5*\nodevsep) {$D$};
  % Row 2: nodes 3, 4
  \node[circle, draw, fill=inferenceColor!40, minimum size=\nodesize, inner sep=0pt] (bm3) at (-0.5*\nodesep, -0.5*\nodevsep) {$M$};
  \node[circle, draw, fill=inferenceColor!40, minimum size=\nodesize, inner sep=0pt] (bm4) at (0.5*\nodesep, -0.5*\nodevsep) {$M$};

  \draw[node arrow] (bm1.east) -- (bm2.west);
  \draw[node arrow] (bm4.north) -- (bm2.south);
  \draw[node arrow] (bm3.north) -- (bm1.south);
\end{scope}

% Bottom-right panel
\draw[line width=\panellinewidth, draw=\panelbordercolor, fill=\panelfillcolor, rounded corners=\panelrounded] 
  (2*\panelwidthsmall + 2*\panelhsep, -\panelheight - \panelvsep) 
  rectangle ++(\panelwidthsmall, \panelheight);

% Bottom-right text box
\begin{scope}[shift={(2*\panelwidthsmall + 2*\panelhsep + 0.5*\panelwidthsmall + \brTextXOffset, -\panelvsep - \brTextYOffset)}]
  \node[anchor=north, text width=\textBoxWidth*\panelwidthsmall, align=center] (brtitle) at (0,0) {\titleFontSize\bfseries\brTitle};
  \node[anchor=north, text width=\textBoxWidth*\panelwidthsmall, align=center] (brdesc) at (0,-0.4cm) {\descFontSize\brDesc};
\end{scope}

% Draw 2×2 grid in bottom-right panel with position offsets
\begin{scope}[shift={(2*\panelwidthsmall + 2*\panelhsep + 0.5*\panelwidthsmall + \brGraphXOffset, -\panelheight - \panelvsep + 0.5*\panelheight + \brGraphYOffset)}]
  % Row 1: nodes 1, 2
  \node[circle, draw, fill=applicationColor!40, minimum size=\nodesize, inner sep=0pt] (br1) at (-0.5*\nodesep, 0.5*\nodevsep) {$D$};
  \node[circle, draw, fill=applicationColor!40, minimum size=\nodesize, inner sep=0pt] (br2) at (0.5*\nodesep, 0.5*\nodevsep) {$D$};
  % Row 2: nodes 3, 4
  \node[circle, draw, fill=applicationColor!40, minimum size=\nodesize, inner sep=0pt] (br3) at (-0.5*\nodesep, -0.5*\nodevsep) {$M$};
  \node[circle, draw, fill=applicationColor!40, minimum size=\nodesize, inner sep=0pt] (br4) at (0.5*\nodesep, -0.5*\nodevsep) {$M$};

  \draw[node arrow] (br1.east) -- (br2.west);
  \draw[node arrow] (br4.north) -- (br2.south);
  \draw[node arrow] (br3.north east) -- (br2.south west);
\end{scope}

% Dashed connections between panels
\draw[node arrow, dashed] (tl3.east) -- (tr1.west);
\draw[node arrow, dashed] (tl6.east) -- (tr4.west);

% tr3 -> tl1  (inverted from tl1 -> tr3)
\draw[node arrow, dashed, black]
  (tr3.north) |- ([yshift=0.9cm]tl1.north) -- (tl1.north);

% bl2 -> tr4  (previously tr4 -> bl2)
\draw[dashed, black] (bl2.north) |- ([yshift=-0.7cm]tr4.south) -- (tr4.south);

% bm2 -> tr5  (previously tr5 -> bm2)
\draw[dashed, black] (bm2.north) |- ([yshift=-0.9cm]tr5.south) -- (tr5.south);

% tr6 -> br2 (go down, then right, then down)
% \draw[node arrow, dashed, black] (tr6.south) -- ([yshift=-0.8cm]tr6.south) -| (br2.north);
\draw[dashed] (br2.north) -- (tr6.south);

% ============================================
% LABELS (drawn last so they appear above arrows)
% ============================================

% --- Top-left: Environment View (E) ---
\node[nodelabel, anchor=north] at ($(tl1.south)+(0,-\nodelabeloffset)$) {Promotion Received};
\node[nodelabel, anchor=north] at ($(tl2.south)+(0,-\nodelabeloffset)$) {Churn (at t+1)};
\node[nodelabel, anchor=north] at ($(tl3.south)+(0,-\nodelabeloffset)$) {User Demographics};
\node[nodelabel, anchor=north] at ($(tl4.south)+(0,-\nodelabeloffset)$) {Economic Factors};
\node[nodelabel, anchor=north] at ($(tl5.south)+(0,-\nodelabeloffset)$) {Quality of Service};
\node[nodelabel, anchor=north] at ($(tl6.south)+(0,-\nodelabeloffset)$) {User Activity};
% --- Top-right: ML System View (O) ---
\node[nodelabel, anchor=north] at ($(tr1.south)+(0,-\nodelabeloffset)$) {Demographic Features};
\node[nodelabel, anchor=north] at ($(tr2.south)+(0,-\nodelabeloffset)$) {Ranked Promotions};
\node[nodelabel, anchor=north] at ($(tr3.south)+(0,-\nodelabeloffset)$) {Promotion Decision};
\node[nodelabel, anchor=north] at ($(tr4.south)+(0,-\nodelabeloffset)$) {Activity Features};
\node[nodelabel, anchor=north] at ($(tr5.south)+(0,-\nodelabeloffset)$) {Churn Risk Score};
\node[nodelabel, anchor=north] at ($(tr6.south)+(0,-\nodelabeloffset)$) {Outreach Decision};
% --- Bottom-left: Data Pipeline View (mix O/C) ---
\node[nodelabel, anchor=north] at ($(bl1.south)+(0,-\nodelabeloffset)$) {Daily Activity};
\node[nodelabel, anchor=north] at ($(bl2.south)+(0,-\nodelabeloffset)$) {Activity Features};
\node[nodelabel, anchor=north] at ($(bl3.south)+(0,-\nodelabeloffset)$) {Parsing Errors};
\node[nodelabel, anchor=north] at ($(bl4.south)+(0,-\nodelabeloffset)$) {Data Freshness};
% --- Bottom-middle: Model Serving / Inference Pipeline (mix O/C) ---
\node[nodelabel, anchor=north] at ($(bm1.south)+(0,-\nodelabeloffset)$) {Loaded Features};
\node[nodelabel, anchor=north] at ($(bm2.south)+(0,-\nodelabeloffset)$) {Churn Risk Score};
\node[nodelabel, anchor=north] at ($(bm3.south)+(0,-\nodelabeloffset)$) {Fetch Timeouts};
\node[nodelabel, anchor=north] at ($(bm4.south)+(0,-\nodelabeloffset)$) {Model Version};
% --- Bottom-right: Application View (mix O/C) ---
\node[nodelabel, anchor=north] at ($(br1.south)+(0,-\nodelabeloffset)$) {Risk Above Threshold};
\node[nodelabel, anchor=north] at ($(br2.south)+(0,-\nodelabeloffset)$) {Outreach Decision};
\node[nodelabel, anchor=north] at ($(br3.south)+(0,-\nodelabeloffset)$) {Burn Rate};
\node[nodelabel, anchor=north] at ($(br4.south)+(0,-\nodelabeloffset)$) {Customer Priority};

\end{tikzpicture}

\vspace{-1.2cm} % Add some space between line and figure
\noindent\rule{\textwidth}{0.4pt}
\vspace{-0.6cm} % Add some space between line and figure
\caption{Illustrative ML System Map composed of five views: \textit{Environment}, \textit{ML System}, and three \textit{Subsystem views} (\textit{data pipeline} for activity features, \textit{model serving} for churn, \textit{application} for outreach). Nodes represent variables: data (D), modulators (M), and random (R). Edges denote relations: causal (solid arrows), mapping (dashed lines), and causal mapping (dashed arrows).}
%We only illustrate three subsystem views, one could imagine there is subsystem view for each node within the ML system view.

% this graph iscaptured on the level of a single observation

\label{fig:system_map}
\end{figure*}

\subsection{Notation}\label{notation}

The notation defines variable types for system quantities and their relations, forming the vocabulary used in our method:

\begin{itemize}
    \item \textbf{Data variables} are populated within the system through managed processes. They represent outputs produced by each step of the data flow -- e.g., features to predictions.
    
    \item \textbf{Modulator variables} condition how output data are produced from their input data -- including configurations (e.g., model version), and operational status (e.g., data freshness). 
    
    \item \textbf{Random variables} denote real-world phenomena that the system observes and/or influences. They remain conceptually distinct from data variables and are not directly observed, but accessed through proxies -- e.g., user activity is observed through activity features.
    
    \item \textbf{Causal relations} denote functional determination between variables. They occur as \emph{data to data} (e.g., features to predictions), \emph{modulator to data} (e.g., model version to predictions), and \emph{random to random} (e.g., user demographics to activity).

\item\textbf{Mapping relations} denote that two data variables represent the same quantity across views (explained in Section \ref{views}) -- e.g., the churn risk score (ML System View) maps to the terminal data variable (Model Serving View)

\item\textbf{Causal mapping relations} connect what the environment produces (random variables) with what the internal system computes (data variables). Unlike mapping relations -- which align quantities across views -- they assert a cross-view causal link and treat the data variable as a proxy for a random variable. In the \emph{random to data} direction, an environmental quantity is captured and processed into an observed data variable that serves as its proxy (e.g., user activity to activity features). In the \emph{data to random} direction, a system data variable brings about an environmental outcome (e.g., outreach decision to promotion received). The relation is not identity -- the two nodes are distinct, linked by measurement or actuation -- and it does not assert independence -- any upstream dependence on the source side (e.g., user demographics influencing activity) remains. 
\end{itemize}
% This keeps data and random variables separate while allowing for a single causal story within each view -- the internal ML system and environment.

% The linked nodes are not identical -- the measurement or actuation pipeline stands between them -- and the mapping does not sever upstream dependencies: any causes of the source variable (e.g., user demographics influencing activity) remain relevant when reasoning about the target. This keeps outputs and random variables distinct while allowing a single causal story within each view -- computation in the system and generation in the environment.

\subsection{Views}\label{views}

We compose variables and relations into three types of views that systematically address the three attribution questions: the \textit{ML System View} (AQ1), \textit{Subsystem views} (AQ2), and the \textit{Environment View} (AQ3). Each data variable in the ML System View represents the output of a subsystem that could be expanded into its own Subsystem view (Figure \ref{fig:system_map} shows three such views). The views follow two principles: \textit{sparsity} -- each view shows only the variables and relations necessary to answer its specific attribution question, focusing attention on relevant factors and also making causal inference tractable \cite{budhathokik2021why, spirtes2010introduction}; and \textit{causal consistency} -- causal relationships are preserved both within each view and when tracing across views.

The \textbf{ML System View} contains only the terminal data variables from each internal subsystem -- the contribution of a subsystem, and thus team, to the ML system's function. This view abstracts away all intermediate data and modulator variables within the subsystems, capturing only the high-level dependencies between them. When we observe drift in a downstream variable, such as final outreach decisions, this sparse structure allows us to identify which upstream subsystem may be responsible (\textbf{AQ1)}.

A \textbf{Subsystem View} exposes the internal structure of a specific subsystem. Each such view includes all intermediate data variables produced along the subsystem's processing pipeline, from initial inputs to the terminal data variables. For example, the Data Pipeline View shows how raw activity counts are transformed into final activity features, with modulator variables for parsing quality and data freshness governing the transformation steps. The terminal data variable is connected via direct mapping to the ML System View above. This structure enables locating which specific component within a subsystem introduced a fault (\textbf{AQ2}).

The \textbf{Environment View} structures random variables according to causal relationships in the underlying data-generating process -- e.g., how user demographics affect activity. Data variables connect to random variables through causal mappings and enable reasoning about environmental quantities when the proxy is valid. For example, activity features reflect actual user activity only if data quality is maintained; outreach decisions translate to promotions received only if delivery succeeds. The view may also include external variables outside the feature set -- such as those omitted during feature selection \cite{lamsaf2025causality} -- to explain shifts that features cannot account for (e.g., quality of service or economic factors) \cite{leest2025tea}. When proxies are valid and the causal structure is modeled, we can determine whether shifts originate from natural external changes (\textbf{AQ3}).

% Note. In the example we show nodes, each representing the output of a team. Data pipelines may chain together, drawing from raw source tables, intermediate datasets, or outputs from prior pipelines. E.g., one team produces raw source data tables. another team has a pipeline that extracts from these tables and produces features stored in a feature store. The running example is kept simple, there are no separate teams shown that produce the source data tables from which features are computed. 

\section{Explaining Distribution Shifts} \label{explaining_drifts}

% We explain how ML System Maps attribute shifts: first defining core concepts, then describing view traversal.

We now turn to the use of \textit{ML System Maps} for attributing shifts. We first define core concepts and then describe view traversal.

\subsection{Attribution and Shift Propagation}\label{attribution_problem}

A distribution shift is a change in the properties of data between current and reference observation windows (e.g., this week vs. last week), measured through divergence metrics \cite{huyen2022}. The attribution problem is to identify the source of the shift \cite{budhathokik2021why}. This requires understanding which variable(s)' mechanisms changed.

The intuition behind attribution follows from the causal graph, where each variable is generated through a \emph{mechanism}. \textbf{Marginal mechanisms} are the distributions $P(X_j)$ of root variables (those with no parents). For example, $P(\text{Demographics})$ captures how demographic characteristics vary in the user population. \textbf{Causal mechanisms} are the conditional distributions $P(X_j|PA_j)$ that define how a variable depends on its direct causes $PA_j$. For example, $P(\text{Activity}|\text{Demographics})$ captures how user activity depends on demographics through some underlying mechanism.

When a variable shifts, it must be explained by a change in one of these mechanisms. A \textbf{change in marginal mechanism} occurs when a root variable's distribution changes and propagates through stable causal mechanisms. For example, if activity shifts while demographics also shift, and $P(\text{Activity}|\text{Demographics})$ remains stable, the activity shift is explained by the demographic change (e.g., we are just observing more young users) propagating through an unchanged mechanism, tracing back to $P(\text{Demographics})$.

A \textbf{change in causal mechanism} explains a shift when the conditional $P(X_j|PA_j)$ itself has changed -- intuitively, this occurs when a variable changes despite its direct causes remaining stable. For example, activity features shifting but demographics remain stable, indicating that some hidden confounding factor caused the shift \cite{vzliobaite2015overview}. Similarly, computational changes are causal mechanism changes -- e.g., if an aggregation job fails and produces NULLs, the mechanism transforming raw data into features has changed \cite{paleyes2023dataflow}.

While ML System Maps remain agnostic to the attribution algorithm used, various attribution methods \cite{singal2021flow, budhathokik2021why, paleyes2023dataflow, cobbo2022contextaware} can identify which mechanism is responsible for a shift. For example, attribution can be performed with Shapley values, as in \cite{budhathokik2021why}. Their method models the mechanisms -- marginal and causal, approximated using empirical and conditional models, respectively. It then swaps mechanisms between windows: for a subset of variables, it uses their current mechanisms while keeping others from the reference window, calculating each mechanism's contribution to the shift. High scores on a variable indicate that its mechanism changed -- marginal (for root variables) or causal (for variables with parents).

% The intuition behind attribution follows from this structure: when we observe a shift downstream, we check whether its parents shifted consistently with their mechanism. If yes, we trace upstream to the origin (a marginal mechanism change). If not, we've found a causal mechanism change at that location, indicating a fault or configuration change that requires investigation.

\subsection{Attribution over ML System Maps}\label{workflow}

When a shift alert fires, we traverse the ML System Map to trace the shift from symptom to source, answering one AQ per view. Within each view, we identify attribution patterns (APs) -- outcomes where attribution mass (e.g., Shapley values) concentrates on specific variables. Figure \ref{fig:workflow} illustrates this process with scenarios where a promotion ranking shift traces to different explanations (for brevity, only the Data Pipeline View is shown for AQ2; other subsystem views follow analogously). We proceed as follows:

\subfile{flow_mappings.tex}
% fig:workflow

\paragraph{\textbf{AQ1 -- \textit{Route}}}
Alerts are defined over \textit{data variables} in \textit{the ML System View}. Routing attributes to the subsystem whose \textit{terminal} data variable best explains a shift; it answers \textit{where}, not \textit{why}. For example, when the ML team managing the promotion-ranking model observes a shift in the rankings, routing on the \textit{ML System View} potentially assigns responsibility upstream. Two APs apply:

\begin{itemize}
\item[\textbf{AP1.1}] \textbf{Subsystem Isolated.}
Mass concentrates on a \textit{conditional data variable}. Upstream data feeding this variable is stable, so the shift is local to its subsystem. Route to the corresponding subsystem (\textit{AQ2}) -- e.g., mass on outreach decisions isolates the outreach application, a policy change might have exposed the model to a different user population.

\item[\textbf{AP1.2}] \textbf{Isolated at Boundary.}
Mass concentrates on a \textit{root data variable} (i.e., features), not a conditional. We first route to the corresponding subsystem that produced the data (\textit{AQ2}). However, the cause may also be in the environment (\textit{AQ3}) -- e.g., mass on activity features can reflect pipeline issues, schema or logic changes, or a natural behavioral shift.
\end{itemize}

\paragraph{\textbf{AQ2 -- \textit{Localize}}}
We open the implicated \textit{Subsystem View} to localize the possible cause of the shift. For example, if mass concentrated on activity features (\textit{AP1.2}), we open the \textit{Data Pipeline View} and attribute to potential faults. The following APs apply:
\vspace{-0.11cm}
\begin{itemize}
\item[\textbf{AP2.1}] \textbf{Root Cause Localized.}
Mass concentrating on a \textit{modulator} indicates the shift was caused by a change in how a function within the subsystem produces data -- e.g., attribution pointing to parsing errors that affected the activity events prior to computing user activity features. The root cause is identified and the procedure terminates.

\item[\textbf{AP2.2}] \textbf{Component Localized.} 
Mass concentrates on an internal \textit{data variable} while no modulator carries mass. The shift is internal to the subsystem -- e.g., mass on activity features indicates failed jobs or logic changes, but manual investigation is required to identify the root cause.

\item[\textbf{AP2.3}] \textbf{Localized at Boundary.} 
Mass concentrates on the \textit{first data variable} at the \textit{data pipeline's} input boundary (e.g., daily activity counts). This boundary is ambiguous between internal issues and genuine external changes. Proceed to \textit{AQ3} to test if external factors explain the shift.
\end{itemize}

\paragraph{\textbf{AQ3 -- \textit{Externalize}.}}
When attribution mass concentrate at the data pipeline's input boundary (AP2.3), we turn to the \textit{Environment View}. Having ruled out internal causes, we return to the feature-level variable that routes to this pipeline (e.g., user activity features), treating it as a proxy for its corresponding random variable to test whether external factors explain the shift. The following APs apply:

\begin{itemize}
\item[\textbf{AP3.1}] \textbf{Explained Externally.} Mass concentrates on an \textit{upstream random variable} (e.g., quality of service) consistent with a known causal relation. The shift reflects external change; assess whether it's transient (e.g., a service outage that is already resolved) or requires model retraining if persistent. 

\item[\textbf{AP3.2}] \textbf{Cannot Determine.} Mass concentrates on the implicated random variable itself (e.g., user activity) rather than its parents, indicating the shift cannot be explained by known external factors. This reflects unresolved ambiguity -- the cause may still be internal or hidden confounders in the environment (e.g., recent product feature releases).

% \item[\textbf{AP3.2}] \textbf{External Cause Unknown.} Mass concentrates on the implicated random variable itself (e.g., user activity) rather than its parents. This indicates a mechanism change -- either from a hidden confounder (e.g., an untracked event affecting activity) or from an internal data quality issue. We proceed on two tracks: (1) we search for candidate hidden confounders; if identified, we extend the map and reduce to \textit{AP3.1}; and (2) we search for missing internal modulators within the data pipeline (following \textit{AP2.2}). For example, unexplained mass on user activity itself suggests either hidden confounders (competitor actions, UI changes) or uninstrumented internal issues.
\end{itemize}

\textbf{Note on Attribution.} These patterns assume clear mass concentration. Scores may be distributed across variables -- e.g., due to noise, or concurrent causes. When distributed, scores guide parallel investigations rather than committing to a single path. The approach applies to any data variable in the ML System View, including adding ground truth (e.g., churn labels) for performance change attribution (prediction-label joint distribution shifts).

% \subsection{nuance}
% We keep attribution sparse by testing only the mechanisms shown in a view; to remain causally consistent we evaluate those tests under a dependency-faithful parent measure shared across views. In other words, sparsity organizes the hypothesis set; a common context fixes the expectations.

% We keep each view sparse by swapping only the mechanisms shown in that panel, but we evaluate those swaps under the dependency-faithful parent distribution implied by the underlying model, not a product of marginals. In the ML System view, for example, we explain a shift in outreach decision by testing only feature, model, and ranking mechanisms while drawing paired samples of demographic features and activity features so their upstream dependence is preserved. In the Environment view, we explain changes by testing only the relevant environmental mechanisms -- mix (a marginal change, such as a shift in the demographics composition) versus behavior (a conditional change, such as activity given demographics) -- and we take expectations with respect to the same joint over demographics and activity. Actions are handled via their causal mapping: promotion received is assessed as the environmental counterpart of the system’s decision, so even if features are not drawn in the Environment panel, the evaluation still conditions on the correct parents (including delivery conditions) from the shared joint. In short, the view limits what we test; a shared context fixes how we take expectations.

\section{Discussion} \label{sec:discussion}

\paragraph{\textbf{Organizational Considerations}} While we focused on technical aspects, organizational considerations are equally central. \textbf{First}, we believe system maps can establish a collaborative diagnostic capability by providing a shared structure for attributing shifts across organizational boundaries. \textbf{Second}, \textit{ML system maps} could create natural incentives for participation. Downstream teams facing the most direct exposure to shifts benefit from instrumenting their dependencies to attribute non-local faults; upstream teams similarly benefit from instrumenting their subsystems when shifts are escalated to them. \textbf{Third}, the approach can consolidate knowledge in two ways: understanding of environmental data-generating processes can be shared across modeling teams, while investigations consolidate findings by extending the map where incomplete. While maintaining a \textit{ML System Map} requires coordination (upfront and continuous, since the system may change), we expect this to be offset by improved traceability and faster incident resolution.

\paragraph{\textbf{Implementation Considerations}} 
Our approach assumes extensive observability -- instrumentation and lineage traceability throughout the ML system. It requires decision-level lineage that extends traceability in both \textbf{scope} and \textbf{granularity}. Current lineage tools often \cite{dbt_docs,namaki2020vamsa} focus on data pipelines, whereas we require end-to-end coverage from data ingestion through model serving to decisions. They also operate at table or column level, whereas observation-level lineage remains an open problem \cite{tomingas2016discovering, cai2025hypnos}. Additionally, such traceability presents substantial overhead through complex temporal alignment across asynchronous subsystems. When such traceability is infeasible, alternative attribution strategies can be considered -- e.g., tracking summary statistics at each node and applying anomaly-based attribution over these \cite{budhathoki2022causal}. While the attribution as illustrated might not be feasible without sufficient observability, we believe the notation and view structure provide value both as theoretical framework and implementation target.

\paragraph{\textbf{Related Work}} We situate our work closely to \cite{leest2025tea}, building on their conceptual framework for describing ML systems, the insight of \cite{paleyes2023dataflow} that flow-based architectures enable fault localization through shift attribution, and the attribution algorithm of \cite{budhathokik2021why}. Our work integrates their insights into a holistic approach.

More broadly, causal modeling has been applied to various aspects of ML and software systems. In software quality assurance, causal graphs have been used for fault localization in microservices -- modeling service dependencies and metrics \cite{zhang2022fault, wu2021causal, li2014causal, chen2016causeinfer, couto2012uncovering}. Within ML systems specifically, causal analysis has addressed isolated concerns: analyzing fairness implications of pipeline and model changes \cite{biswas2021fair, ji2023causality, monjezi2024causal}, and attributing performance degradations to external changes \cite{zhangh2023why}. Our approach differs by providing a holistic, system-wide model of operational ML systems that captures both internal computation and external processes, enabling causal attribution of distribution shifts across the entire system.

% \paragraph{\textbf{Research Agenda}} The approach presented is an envisioned systematic method relying heavily on observability. Significant work remains to operationalize \textit{ML System Maps} in practice. 
% \textbf{First}, controlled experiments on simulated systems are needed to validate core concepts and test alternative modeling, representation, and attribution approaches under varying levels of observability.  \textbf{Second}, case studies in production systems are needed to establish practical feasibility -- starting with proof of concepts, incrementally extending the mapping, surfacing organizational barriers, and addressing technical challenges.

\paragraph{\textbf{Research Agenda}} Significant work remains to operationalize \textit{ML System Maps} in practice. \textbf{First}, controlled experiments on simulated systems are needed to validate that the view structure enables systematic attribution under different levels of observability and map completeness. \textbf{Second}, case studies in production systems are needed to establish practical and organizational feasibility by identifying strategies for incremental map construction, sustained cross-team coordination, and addressing technical challenges. \textbf{Third}, across both experimental and production settings, comparative evaluation of causal attribution algorithms \cite{cobbo2022contextaware, budhathokik2021why, paleyes2023dataflow, singal2021flow} is needed to assess performance tradeoffs between attribution accuracy, computational cost, and observability requirements.


% \bibliographystyle{ACM-Reference-Format}
% \bibliography{bibliography}

\begin{thebibliography}{47}

%%% the Web address.

\ifx \showCODEN    \undefined \def \showCODEN     #1{\unskip}     \fi
\ifx \showDOI      \undefined \def \showDOI       #1{#1}\fi
\ifx \showISBNx    \undefined \def \showISBNx     #1{\unskip}     \fi
\ifx \showISBNxiii \undefined \def \showISBNxiii  #1{\unskip}     \fi
\ifx \showISSN     \undefined \def \showISSN      #1{\unskip}     \fi
\ifx \showLCCN     \undefined \def \showLCCN      #1{\unskip}     \fi
\ifx \shownote     \undefined \def \shownote      #1{#1}          \fi
\ifx \showarticletitle \undefined \def \showarticletitle #1{#1}   \fi
\ifx \showURL      \undefined \def \showURL       {\relax}        \fi
% invisible to TeX
\providecommand\bibfield[2]{#2}
\providecommand\bibinfo[2]{#2}
\providecommand\natexlab[1]{#1}
\providecommand\showeprint[2][]{arXiv:#2}

\bibitem[Biswas and Rajan(2021)]%
        {biswas2021fair}
\bibfield{author}{\bibinfo{person}{Sumon Biswas} {and} \bibinfo{person}{Hridesh Rajan}.} \bibinfo{year}{2021}\natexlab{}.
\newblock \showarticletitle{Fair preprocessing: towards understanding compositional fairness of data transformers in machine learning pipeline}. In \bibinfo{booktitle}{\emph{Proceedings of the 29th ACM Joint Meeting on European Software Engineering Conference and Symposium on the Foundations of Software Engineering}}. \bibinfo{pages}{981--993}.
\newblock


\bibitem[Breck et~al\mbox{.}(2017)]%
        {breck2017ml}
\bibfield{author}{\bibinfo{person}{Eric Breck}, \bibinfo{person}{Shanqing Cai}, \bibinfo{person}{Eric Nielsen}, \bibinfo{person}{Michael Salib}, {and} \bibinfo{person}{D Sculley}.} \bibinfo{year}{2017}\natexlab{}.
\newblock \showarticletitle{The ML test score: A rubric for ML production readiness and technical debt reduction}. In \bibinfo{booktitle}{\emph{2017 IEEE International Conference on Big Data (Big Data)}}. IEEE, \bibinfo{pages}{1123--1132}.
\newblock


\bibitem[Breck et~al\mbox{.}(2019)]%
        {breck2019data}
\bibfield{author}{\bibinfo{person}{Eric Breck}, \bibinfo{person}{Neoklis Polyzotis}, \bibinfo{person}{Sudip Roy}, \bibinfo{person}{Steven Whang}, {and} \bibinfo{person}{Martin Zinkevich}.} \bibinfo{year}{2019}\natexlab{}.
\newblock \showarticletitle{Data Validation for Machine Learning.}. In \bibinfo{booktitle}{\emph{MLSys}}.
\newblock


\bibitem[Budhathoki et~al\mbox{.}(2022)]%
        {budhathoki2022causal}
\bibfield{author}{\bibinfo{person}{Kailash Budhathoki}, \bibinfo{person}{Lenon Minorics}, \bibinfo{person}{Patrick Bl{\"o}baum}, {and} \bibinfo{person}{Dominik Janzing}.} \bibinfo{year}{2022}\natexlab{}.
\newblock \showarticletitle{Causal structure-based root cause analysis of outliers}. In \bibinfo{booktitle}{\emph{International conference on machine learning}}. PMLR, \bibinfo{pages}{2357--2369}.
\newblock


\bibitem[Cai et~al\mbox{.}(2025)]%
        {cai2025hypnos}
\bibfield{author}{\bibinfo{person}{Xiwen Cai}, \bibinfo{person}{Xiaodong Ge}, \bibinfo{person}{Kai Xiong}, \bibinfo{person}{Shuainan Ye}, \bibinfo{person}{Di Weng}, \bibinfo{person}{Ke Xu}, \bibinfo{person}{Datong Wei}, \bibinfo{person}{Jiang Long}, {and} \bibinfo{person}{Yingcai Wu}.} \bibinfo{year}{2025}\natexlab{}.
\newblock \showarticletitle{HYPNOS: Interactive Data Lineage Tracing for Data Transformation Scripts}.
\newblock \bibinfo{journal}{\emph{IEEE Transactions on Visualization and Computer Graphics}} (\bibinfo{year}{2025}).
\newblock


\bibitem[Chen et~al\mbox{.}(2016)]%
        {chen2016causeinfer}
\bibfield{author}{\bibinfo{person}{Pengfei Chen}, \bibinfo{person}{Yong Qi}, {and} \bibinfo{person}{Di Hou}.} \bibinfo{year}{2016}\natexlab{}.
\newblock \showarticletitle{CauseInfer: Automated end-to-end performance diagnosis with hierarchical causality graph in cloud environment}.
\newblock \bibinfo{journal}{\emph{IEEE transactions on services computing}} \bibinfo{volume}{12}, \bibinfo{number}{2} (\bibinfo{year}{2016}), \bibinfo{pages}{214--230}.
\newblock


\bibitem[Cobb and Van~Looveren(2022)]%
        {cobbo2022contextaware}
\bibfield{author}{\bibinfo{person}{Oliver Cobb} {and} \bibinfo{person}{Arnaud Van~Looveren}.} \bibinfo{year}{2022}\natexlab{}.
\newblock \showarticletitle{Context-aware drift detection}. In \bibinfo{booktitle}{\emph{International conference on machine learning}}. PMLR, \bibinfo{pages}{4087--4111}.
\newblock


\bibitem[Couto et~al\mbox{.}(2012)]%
        {couto2012uncovering}
\bibfield{author}{\bibinfo{person}{Cesar Couto}, \bibinfo{person}{Christofer Silva}, \bibinfo{person}{Marco~Tulio Valente}, \bibinfo{person}{Roberto Bigonha}, {and} \bibinfo{person}{Nicolas Anquetil}.} \bibinfo{year}{2012}\natexlab{}.
\newblock \showarticletitle{Uncovering causal relationships between software metrics and bugs}. In \bibinfo{booktitle}{\emph{2012 16th European Conference on Software Maintenance and Reengineering}}. IEEE, \bibinfo{pages}{223--232}.
\newblock


\bibitem[{dbt Labs}(2025)]%
        {dbt_docs}
\bibfield{author}{\bibinfo{person}{{dbt Labs}}.} \bibinfo{year}{2025}\natexlab{}.
\newblock \bibinfo{title}{dbt Developer Hub}.
\newblock \bibinfo{howpublished}{\url{https://docs.getdbt.com/}}.
\newblock
\newblock
\shownote{Documentation site. Accessed 2025-10-17}.


\bibitem[Ehrlinger et~al\mbox{.}(2019)]%
        {ehrlingerl2019a}
\bibfield{author}{\bibinfo{person}{Lisa Ehrlinger}, \bibinfo{person}{Verena Haunschmid}, \bibinfo{person}{Davide Palazzini}, {and} \bibinfo{person}{Christian Lettner}.} \bibinfo{year}{2019}\natexlab{}.
\newblock \showarticletitle{A DaQL to monitor data quality in machine learning applications}.
\newblock  (\bibinfo{year}{2019}), \bibinfo{pages}{227--237}.
\newblock


\bibitem[Feng et~al\mbox{.}(2024a)]%
        {feng2024monitoring}
\bibfield{author}{\bibinfo{person}{Jean Feng}, \bibinfo{person}{Alexej Gossmann}, \bibinfo{person}{Gene~A Pennello}, \bibinfo{person}{Nicholas Petrick}, \bibinfo{person}{Berkman Sahiner}, {and} \bibinfo{person}{Romain Pirracchio}.} \bibinfo{year}{2024}\natexlab{a}.
\newblock \showarticletitle{Monitoring machine learning-based risk prediction algorithms in the presence of performativity}. In \bibinfo{booktitle}{\emph{International Conference on Artificial Intelligence and Statistics}}. PMLR, \bibinfo{pages}{919--927}.
\newblock


\bibitem[Feng et~al\mbox{.}(2024b)]%
        {fengj2024designing}
\bibfield{author}{\bibinfo{person}{Jean Feng}, \bibinfo{person}{Adarsh Subbaswamy}, \bibinfo{person}{Alexej Gossmann}, \bibinfo{person}{Harvineet Singh}, \bibinfo{person}{Berkman Sahiner}, \bibinfo{person}{Mi-Ok Kim}, \bibinfo{person}{Gene~Anthony Pennello}, \bibinfo{person}{Nicholas Petrick}, \bibinfo{person}{Romain Pirracchio}, {and} \bibinfo{person}{Fan Xia}.} \bibinfo{year}{2024}\natexlab{b}.
\newblock \showarticletitle{Designing monitoring strategies for deployed machine learning algorithms: navigating performativity through a causal lens}. In \bibinfo{booktitle}{\emph{Causal Learning and Reasoning}}. PMLR, \bibinfo{pages}{587--608}.
\newblock


\bibitem[Google(2024)]%
        {tfx}
\bibfield{author}{\bibinfo{person}{Google}.} \bibinfo{year}{2024}\natexlab{}.
\newblock \bibinfo{title}{TensorFlow Extended (TFX)}.
\newblock
\newblock
\urldef\tempurl%
\url{https://www.tensorflow.org/tfx}
\showURL{%
\tempurl}


\bibitem[Huyen(2022)]%
        {huyen2022}
\bibfield{author}{\bibinfo{person}{Chip Huyen}.} \bibinfo{year}{2022}\natexlab{}.
\newblock \bibinfo{booktitle}{\emph{Designing machine learning systems}}.
\newblock \bibinfo{publisher}{" O'Reilly Media, Inc."}.
\newblock


\bibitem[Ji et~al\mbox{.}(2023)]%
        {ji2023causality}
\bibfield{author}{\bibinfo{person}{Zhenlan Ji}, \bibinfo{person}{Pingchuan Ma}, \bibinfo{person}{Shuai Wang}, {and} \bibinfo{person}{Yanhui Li}.} \bibinfo{year}{2023}\natexlab{}.
\newblock \showarticletitle{Causality-aided trade-off analysis for machine learning fairness}. In \bibinfo{booktitle}{\emph{2023 38th IEEE/ACM International Conference on Automated Software Engineering (ASE)}}. IEEE, \bibinfo{pages}{371--383}.
\newblock


\bibitem[Kailash et~al\mbox{.}(2021)]%
        {budhathokik2021why}
\bibfield{author}{\bibinfo{person}{Budhathoki Kailash}, \bibinfo{person}{Janzing Dominik}, \bibinfo{person}{Blöbaum Patrick}, {and} \bibinfo{person}{Ng Hoiyi}.} \bibinfo{year}{2021}\natexlab{}.
\newblock \showarticletitle{Why did the distribution change?}. In \bibinfo{booktitle}{\emph{Proceedings of Machine Learning Research}}.
\newblock


\bibitem[Kreuzberger et~al\mbox{.}(2023)]%
        {kreuzberger2023machine}
\bibfield{author}{\bibinfo{person}{Dominik Kreuzberger}, \bibinfo{person}{Niklas K{\"u}hl}, {and} \bibinfo{person}{Sebastian Hirschl}.} \bibinfo{year}{2023}\natexlab{}.
\newblock \showarticletitle{Machine learning operations (mlops): Overview, definition, and architecture}.
\newblock \bibinfo{journal}{\emph{IEEE Access}} (\bibinfo{year}{2023}).
\newblock


\bibitem[Labs(2021)]%
        {cloudera2021}
\bibfield{author}{\bibinfo{person}{Cloudera Fast~Forward Labs}.} \bibinfo{year}{2021}\natexlab{}.
\newblock \showarticletitle{Inferring Concept Drift Without Labeled Data}.
\newblock  (\bibinfo{year}{2021}).
\newblock


\bibitem[Lamsaf et~al\mbox{.}(2025)]%
        {lamsaf2025causality}
\bibfield{author}{\bibinfo{person}{Asmae Lamsaf}, \bibinfo{person}{Rui Carrilho}, \bibinfo{person}{Jo{\~a}o~C Neves}, {and} \bibinfo{person}{Hugo Proen{\c{c}}a}.} \bibinfo{year}{2025}\natexlab{}.
\newblock \showarticletitle{Causality, machine learning, and feature selection: a survey}.
\newblock \bibinfo{journal}{\emph{Sensors}} \bibinfo{volume}{25}, \bibinfo{number}{8} (\bibinfo{year}{2025}), \bibinfo{pages}{2373}.
\newblock


\bibitem[Leest et~al\mbox{.}(2025)]%
        {leest2025tea}
\bibfield{author}{\bibinfo{person}{Joran Leest}, \bibinfo{person}{Claudia Raibulet}, \bibinfo{person}{Patricia Lago}, {and} \bibinfo{person}{Ilias Gerostathopoulos}.} \bibinfo{year}{2025}\natexlab{}.
\newblock \showarticletitle{From Tea Leaves to System Maps: A Survey and Framework on Context-aware Machine Learning Monitoring}.
\newblock \bibinfo{journal}{\emph{IEEE Transactions on Software Engineering}} (\bibinfo{year}{2025}).
\newblock


\bibitem[Li et~al\mbox{.}(2014)]%
        {li2014causal}
\bibfield{author}{\bibinfo{person}{Lixian Li}, \bibinfo{person}{Jin Liu}, \bibinfo{person}{Zhangbing Zhou}, \bibinfo{person}{Haoyu Luo}, \bibinfo{person}{Wenrui Liu}, {and} \bibinfo{person}{Juan Li}.} \bibinfo{year}{2014}\natexlab{}.
\newblock \showarticletitle{Causal inference based service dependency graph for statistical service fault localization}. In \bibinfo{booktitle}{\emph{2014 10th International Conference on Semantics, Knowledge and Grids}}. IEEE, \bibinfo{pages}{41--48}.
\newblock


\bibitem[Lu et~al\mbox{.}(2019)]%
        {lu2018}
\bibfield{author}{\bibinfo{person}{Jie Lu}, \bibinfo{person}{Anjin Liu}, \bibinfo{person}{Fan Dong}, \bibinfo{person}{Feng Gu}, \bibinfo{person}{Joao Gama}, {and} \bibinfo{person}{Guangquan Zhang}.} \bibinfo{year}{2019}\natexlab{}.
\newblock \showarticletitle{Learning under concept drift: A review}.
\newblock \bibinfo{journal}{\emph{IEEE Transactions on Knowledge and Data Engineering}} \bibinfo{volume}{31}, \bibinfo{number}{12} (\bibinfo{year}{2019}), \bibinfo{pages}{2346--2363}.
\newblock
\urldef\tempurl%
\url{https://doi.org/10.1109/TKDE.2018.2876857}
\showDOI{\tempurl}


\bibitem[Monjezi et~al\mbox{.}(2024)]%
        {monjezi2024causal}
\bibfield{author}{\bibinfo{person}{Verya Monjezi}, \bibinfo{person}{Ashish Kumar}, \bibinfo{person}{Gang Tan}, \bibinfo{person}{Ashutosh Trivedi}, {and} \bibinfo{person}{Saeid Tizpaz-Niari}.} \bibinfo{year}{2024}\natexlab{}.
\newblock \showarticletitle{Causal graph fuzzing for fair ML sofware development}. In \bibinfo{booktitle}{\emph{Proceedings of the 2024 IEEE/ACM 46th International Conference on Software Engineering: Companion Proceedings}}.
\newblock


\bibitem[Namaki et~al\mbox{.}(2020)]%
        {namaki2020vamsa}
\bibfield{author}{\bibinfo{person}{Mohammad~Hossein Namaki}, \bibinfo{person}{Avrilia Floratou}, \bibinfo{person}{Fotis Psallidas}, \bibinfo{person}{Subru Krishnan}, \bibinfo{person}{Ashvin Agrawal}, \bibinfo{person}{Yinghui Wu}, \bibinfo{person}{Yiwen Zhu}, {and} \bibinfo{person}{Markus Weimer}.} \bibinfo{year}{2020}\natexlab{}.
\newblock \showarticletitle{Vamsa: Automated provenance tracking in data science scripts}. In \bibinfo{booktitle}{\emph{Proceedings of the 26th ACM SIGKDD international conference on knowledge discovery \& data mining}}. \bibinfo{pages}{1542--1551}.
\newblock


\bibitem[Paleyes et~al\mbox{.}(2023)]%
        {paleyes2023dataflow}
\bibfield{author}{\bibinfo{person}{A Paleyes}, \bibinfo{person}{Siyuan Guo}, \bibinfo{person}{Bernhard Sch{\"o}lkopf}, {and} \bibinfo{person}{ND Lawrence}.} \bibinfo{year}{2023}\natexlab{}.
\newblock \showarticletitle{Dataflow graphs as complete causal graphs}. In \bibinfo{booktitle}{\emph{IEEE/ACM 2nd International Conference on AI Engineering--Software Engineering for AI (CAIN 2023)}}. IEEE, \bibinfo{pages}{7--12}.
\newblock


\bibitem[Protschky et~al\mbox{.}(2025)]%
        {protschky2025gets}
\bibfield{author}{\bibinfo{person}{Dominik Protschky}, \bibinfo{person}{Luis L{\"a}mmermann}, \bibinfo{person}{Peter Hofmann}, {and} \bibinfo{person}{Nils Urbach}.} \bibinfo{year}{2025}\natexlab{}.
\newblock \showarticletitle{What Gets Measured Gets Improved: Monitoring Machine Learning Applications in Their Production Environments}.
\newblock \bibinfo{journal}{\emph{IEEE Access}} (\bibinfo{year}{2025}).
\newblock


\bibitem[Quintas-Martinez et~al\mbox{.}(2024)]%
        {quintas2024multiply}
\bibfield{author}{\bibinfo{person}{Victor Quintas-Martinez}, \bibinfo{person}{Mohammad~Taha Bahadori}, \bibinfo{person}{Eduardo Santiago}, \bibinfo{person}{Jeff Mu}, \bibinfo{person}{Dominik Janzing}, {and} \bibinfo{person}{David Heckerman}.} \bibinfo{year}{2024}\natexlab{}.
\newblock \showarticletitle{Multiply-robust causal change attribution}.
\newblock \bibinfo{journal}{\emph{arXiv preprint arXiv:2404.08839}} (\bibinfo{year}{2024}).
\newblock


\bibitem[Rudin et~al\mbox{.}(2022)]%
        {rudin2022interpretable}
\bibfield{author}{\bibinfo{person}{Cynthia Rudin}, \bibinfo{person}{Chaofan Chen}, \bibinfo{person}{Zhi Chen}, \bibinfo{person}{Haiyang Huang}, \bibinfo{person}{Lesia Semenova}, {and} \bibinfo{person}{Chudi Zhong}.} \bibinfo{year}{2022}\natexlab{}.
\newblock \showarticletitle{Interpretable machine learning: Fundamental principles and 10 grand challenges}.
\newblock \bibinfo{journal}{\emph{Statistic Surveys}}  \bibinfo{volume}{16} (\bibinfo{year}{2022}), \bibinfo{pages}{1--85}.
\newblock


\bibitem[Salesforce({[n.\,d.]})]%
        {salesforce_marketing_automation}
\bibfield{author}{\bibinfo{person}{Salesforce}.} \bibinfo{year}{[n.\,d.]}\natexlab{}.
\newblock \bibinfo{title}{Marketing Automation Software | Salesforce}.
\newblock \bibinfo{howpublished}{\url{https://www.salesforce.com/eu/marketing/automation/}}.
\newblock
\newblock
\shownote{Accessed: 2025-10-16}.


\bibitem[Sculley et~al\mbox{.}(2015)]%
        {sculley2015hidden}
\bibfield{author}{\bibinfo{person}{David Sculley}, \bibinfo{person}{Gary Holt}, \bibinfo{person}{Daniel Golovin}, \bibinfo{person}{Eugene Davydov}, \bibinfo{person}{Todd Phillips}, \bibinfo{person}{Dietmar Ebner}, \bibinfo{person}{Vinay Chaudhary}, \bibinfo{person}{Michael Young}, \bibinfo{person}{Jean-Francois Crespo}, {and} \bibinfo{person}{Dan Dennison}.} \bibinfo{year}{2015}\natexlab{}.
\newblock \showarticletitle{Hidden technical debt in machine learning systems}.
\newblock \bibinfo{journal}{\emph{Advances in neural information processing systems}}  \bibinfo{volume}{28} (\bibinfo{year}{2015}).
\newblock


\bibitem[Shankar et~al\mbox{.}(2023)]%
        {shankar2023automatic}
\bibfield{author}{\bibinfo{person}{Shreya Shankar}, \bibinfo{person}{Labib Fawaz}, \bibinfo{person}{Karl Gyllstrom}, {and} \bibinfo{person}{Aditya Parameswaran}.} \bibinfo{year}{2023}\natexlab{}.
\newblock \showarticletitle{Automatic and Precise Data Validation for Machine Learning}. In \bibinfo{booktitle}{\emph{Proceedings of the 32nd ACM International Conference on Information and Knowledge Management}}. \bibinfo{pages}{2198--2207}.
\newblock


\bibitem[Shankar et~al\mbox{.}(2024)]%
        {shankars2024we}
\bibfield{author}{\bibinfo{person}{Shreya Shankar}, \bibinfo{person}{Rolando Garcia}, \bibinfo{person}{Joseph~M Hellerstein}, {and} \bibinfo{person}{Aditya~G Parameswaran}.} \bibinfo{year}{2024}\natexlab{}.
\newblock \showarticletitle{" We Have No Idea How Models will Behave in Production until Production": How Engineers Operationalize Machine Learning}.
\newblock \bibinfo{journal}{\emph{Proceedings of the ACM on Human-Computer Interaction}} \bibinfo{volume}{8}, \bibinfo{number}{CSCW1}, \bibinfo{pages}{1--34}.
\newblock


\bibitem[Shankar and Parameswaran(2022)]%
        {shankars2022towards}
\bibfield{author}{\bibinfo{person}{Shreya Shankar} {and} \bibinfo{person}{Aditya~G Parameswaran}.} \bibinfo{year}{2022}\natexlab{}.
\newblock \showarticletitle{Towards Observability for Machine Learning Pipelines.}. In \bibinfo{booktitle}{\emph{CIDR}}.
\newblock


\bibitem[She et~al\mbox{.}(2025)]%
        {she2025fairsense}
\bibfield{author}{\bibinfo{person}{Yining She}, \bibinfo{person}{Sumon Biswas}, \bibinfo{person}{Christian K{\"a}stner}, {and} \bibinfo{person}{Eunsuk Kang}.} \bibinfo{year}{2025}\natexlab{}.
\newblock \showarticletitle{FairSense: Long-Term Fairness Analysis of ML-Enabled Systems}.
\newblock \bibinfo{journal}{\emph{arXiv preprint arXiv:2501.01665}} (\bibinfo{year}{2025}).
\newblock


\bibitem[Shergadwala et~al\mbox{.}(2022)]%
        {shergadwalamn2022a}
\bibfield{author}{\bibinfo{person}{Murtuza~N Shergadwala}, \bibinfo{person}{Himabindu Lakkaraju}, {and} \bibinfo{person}{Krishnaram Kenthapadi}.} \bibinfo{year}{2022}\natexlab{}.
\newblock \showarticletitle{A human-centric perspective on model monitoring}. In \bibinfo{booktitle}{\emph{Proceedings of the AAAI Conference on Human Computation and Crowdsourcing}}, Vol.~\bibinfo{volume}{10}. \bibinfo{pages}{173--183}.
\newblock


\bibitem[Singal et~al\mbox{.}(2021)]%
        {singal2021flow}
\bibfield{author}{\bibinfo{person}{Raghav Singal}, \bibinfo{person}{George Michailidis}, {and} \bibinfo{person}{Hoiyi Ng}.} \bibinfo{year}{2021}\natexlab{}.
\newblock \showarticletitle{Flow-based attribution in graphical models: A recursive shapley approach}. In \bibinfo{booktitle}{\emph{International Conference on Machine Learning}}. PMLR, \bibinfo{pages}{9733--9743}.
\newblock


\bibitem[Spirtes(2010)]%
        {spirtes2010introduction}
\bibfield{author}{\bibinfo{person}{Peter Spirtes}.} \bibinfo{year}{2010}\natexlab{}.
\newblock \showarticletitle{Introduction to causal inference.}
\newblock \bibinfo{journal}{\emph{Journal of Machine Learning Research}} \bibinfo{volume}{11}, \bibinfo{number}{5} (\bibinfo{year}{2010}).
\newblock


\bibitem[Tomingas et~al\mbox{.}(2016)]%
        {tomingas2016discovering}
\bibfield{author}{\bibinfo{person}{Kalle Tomingas}, \bibinfo{person}{Priit J{\"a}rv}, {and} \bibinfo{person}{Tanel Tammet}.} \bibinfo{year}{2016}\natexlab{}.
\newblock \showarticletitle{Discovering Data Lineage from Data Warehouse Procedures}. In \bibinfo{booktitle}{\emph{International Conference on Knowledge Discovery and Information Retrieval}}, Vol.~\bibinfo{volume}{2}. SCITEPRESS, \bibinfo{pages}{101--110}.
\newblock


\bibitem[Webb et~al\mbox{.}(2016)]%
        {webb2016characterizing}
\bibfield{author}{\bibinfo{person}{Geoffrey~I Webb}, \bibinfo{person}{Roy Hyde}, \bibinfo{person}{Hong Cao}, \bibinfo{person}{Hai~Long Nguyen}, {and} \bibinfo{person}{Francois Petitjean}.} \bibinfo{year}{2016}\natexlab{}.
\newblock \showarticletitle{Characterizing concept drift}.
\newblock \bibinfo{journal}{\emph{Data Mining and Knowledge Discovery}} \bibinfo{volume}{30}, \bibinfo{number}{4} (\bibinfo{year}{2016}), \bibinfo{pages}{964--994}.
\newblock


\bibitem[Wellman and Henrion(1993)]%
        {wellman1993explaining}
\bibfield{author}{\bibinfo{person}{Michael~P Wellman} {and} \bibinfo{person}{Max Henrion}.} \bibinfo{year}{1993}\natexlab{}.
\newblock \showarticletitle{Explaining'explaining away'}.
\newblock \bibinfo{journal}{\emph{IEEE Transactions on Pattern Analysis and Machine Intelligence}} \bibinfo{volume}{15}, \bibinfo{number}{3} (\bibinfo{year}{1993}), \bibinfo{pages}{287--292}.
\newblock


\bibitem[Wu et~al\mbox{.}(2021)]%
        {wu2021causal}
\bibfield{author}{\bibinfo{person}{Li Wu}, \bibinfo{person}{Johan Tordsson}, \bibinfo{person}{Erik Elmroth}, {and} \bibinfo{person}{Odej Kao}.} \bibinfo{year}{2021}\natexlab{}.
\newblock \showarticletitle{Causal inference techniques for microservice performance diagnosis: Evaluation and guiding recommendations}. In \bibinfo{booktitle}{\emph{2021 IEEE International Conference on Autonomic Computing and Self-Organizing Systems (ACSOS)}}. IEEE, \bibinfo{pages}{21--30}.
\newblock


\bibitem[Xu et~al\mbox{.}(2023)]%
        {xuz2023alertiger}
\bibfield{author}{\bibinfo{person}{Zhentao Xu}, \bibinfo{person}{Ruoying Wang}, \bibinfo{person}{Girish Balaji}, \bibinfo{person}{Manas Bundele}, \bibinfo{person}{Xiaofei Liu}, \bibinfo{person}{Leo Liu}, {and} \bibinfo{person}{Tie Wang}.} \bibinfo{year}{2023}\natexlab{}.
\newblock \showarticletitle{Alertiger: Deep learning for ai model health monitoring at linkedin}. In \bibinfo{booktitle}{\emph{Proceedings of the 29th ACM SIGKDD Conference on Knowledge Discovery and Data Mining}}.
\newblock


\bibitem[Zapier({[n.\,d.]})]%
        {zapier_workflows}
\bibfield{author}{\bibinfo{person}{Zapier}.} \bibinfo{year}{[n.\,d.]}\natexlab{}.
\newblock \bibinfo{title}{Build automated workflows with Zapier}.
\newblock \bibinfo{howpublished}{\url{https://zapier.com/workflows}}.
\newblock
\newblock
\shownote{Accessed: 2025-10-16}.


\bibitem[Zhang et~al\mbox{.}(2020)]%
        {zhang2020data}
\bibfield{author}{\bibinfo{person}{Amy~X Zhang}, \bibinfo{person}{Michael Muller}, {and} \bibinfo{person}{Dakuo Wang}.} \bibinfo{year}{2020}\natexlab{}.
\newblock \showarticletitle{How do data science workers collaborate? roles, workflows, and tools}.
\newblock \bibinfo{journal}{\emph{Proceedings of the ACM on Human-Computer Interaction}} \bibinfo{volume}{4}, \bibinfo{number}{CSCW1} (\bibinfo{year}{2020}), \bibinfo{pages}{1--23}.
\newblock


\bibitem[Zhang et~al\mbox{.}(2023)]%
        {zhangh2023why}
\bibfield{author}{\bibinfo{person}{Haoran Zhang}, \bibinfo{person}{Harvineet Singh}, \bibinfo{person}{Marzyeh Ghassemi}, {and} \bibinfo{person}{Shalmali Joshi}.} \bibinfo{year}{2023}\natexlab{}.
\newblock \showarticletitle{“Why did the Model Fail?”: Attributing Model Performance Changes to Distribution Shifts}. In \bibinfo{booktitle}{\emph{Proceedings of Machine Learning Research}}.
\newblock


\bibitem[Zhang et~al\mbox{.}(2022)]%
        {zhang2022fault}
\bibfield{author}{\bibinfo{person}{Qixun Zhang}, \bibinfo{person}{Tong Jia}, \bibinfo{person}{Zhonghai Wu}, \bibinfo{person}{Qingxin Wu}, \bibinfo{person}{Lichun Jia}, \bibinfo{person}{Donglei Li}, \bibinfo{person}{Yuqing Tao}, {and} \bibinfo{person}{Yutong Xiao}.} \bibinfo{year}{2022}\natexlab{}.
\newblock \showarticletitle{Fault localization for microservice applications with system logs and monitoring metrics}. In \bibinfo{booktitle}{\emph{2022 7th International Conference on Cloud Computing and Big Data Analytics (ICCCBDA)}}. IEEE, \bibinfo{pages}{149--154}.
\newblock


\bibitem[Zliobaite et~al\mbox{.}(2015)]%
        {vzliobaite2015overview}
\bibfield{author}{\bibinfo{person}{Indre Zliobaite}, \bibinfo{person}{Mykola Pechenizkiy}, {and} \bibinfo{person}{Joao Gama}.} \bibinfo{year}{2015}\natexlab{}.
\newblock \showarticletitle{An overview of concept drift applications}.
\newblock \bibinfo{journal}{\emph{Big data analysis: new algorithms for a new society}} (\bibinfo{year}{2015}), \bibinfo{pages}{91--114}.
\newblock


\end{thebibliography}

\end{document}